# Realization of cold atom gyroscope in space

Jinting Li[1,2], Xi Chen[1]*, Danfang Zhang[1,2], Wenzhang Wang[1,2], Yang Zhou[1,2], Meng He[1], Jie Fang[1], Lin Zhou[1,3], Chuan He[1], Junjie Jiang[1,2], Huanyao Sun[1], Qunfeng Chen[1], Lei Qin[1], Xiao Li[1], Yibo Wang[1], Xiaowei Zhang[1], Jiaqi Zhong[1,3], Runbing Li[1,3,5], Meizhen An[4], Long Zhang[4], Shuquan Wang[4], Zongfeng Li[4], Jin Wang[1,3,5]† and Mingsheng Zhan[1,3,5]‡

**AFFILIATIONS**
[1]State Key Laboratory of Magnetic Resonance and Atomic and Molecular Physics, Innovation Academy for Precision Measurement Science and Technology, Chinese Academy of Sciences, Wuhan 430071, China
[2]School of Physical Sciences, University of Chinese Academy of Sciences, Beijing 100049, China
[3]Hefei National Laboratory, Hefei 230088, China
[4]Technology and Engineering Center for Space Utilization, Chinese Academy of Sciences, Beijing 100094, China
[5]Wuhan Institute of Quantum Technology, Wuhan 430206, China
*Corresponding author. Email: chenxi@apm.ac.cn
†Corresponding author. Email: wangjin@apm.ac.cn
‡Corresponding author. Email: mszhan@apm.ac.cn

**ABSTRACT**
High-precision gyroscopes in space are essential for fundamental physics research and navigation. Due to its potential high precision, the cold atom gyroscope is expected to be the next generation of gyroscopes in space. Here, we report the first realization of a cold atom gyroscope, which was demonstrated by the atom interferometer installed in the China Space Station (CSS) as a payload. By compensating for CSS's high dynamic rotation rate using a built-in piezoelectric mirror, spatial interference fringes in the interferometer are successfully obtained. Then, the optimized ratio of the Raman laser's angles is derived, the coefficients of the piezoelectric mirror are self-calibrated in orbit, and various systemic effects are corrected. We achieve a rotation measurement resolution of 50 μrad/s for a single shot and 17 μrad/s for an average number of 32. The measured rotation is -1142±29 μrad/s and is compatible with that recorded by the classical gyroscope of the CSS. This study paves the way for developing high-precision cold atom gyroscopes in space.

**INTRODUCTION**
  Space-based gyroscopes are important in inertial navigation and fundamental physics tests [1-5]. One typical example is the Gravity Probe B (GP-B) satellite which utilizes cryogenic gyroscopes to test general relativity. The GP-B project gives a test precision of the frame-dragging effect of 19% with one year of measurement data [3]. No violation was observed for this general relativity effect from its theoretical prediction. Further experiments with the LARES 2 satellite [6] in space and GINGER [7] on the ground are on the way to continuously improve the test precision.

Atom interferometers (AIs) are expected to be next-generation gyroscopes to measure rotation with very high precision, as already demonstrated on the ground [8-11]. In space, AIs could achieve a much longer interference time than the ground, thus forming a space gyroscope with comparable precision to the cryogenic gyroscope. For example, the Hyper project is designed to have a rotation measurement resolution of $10^{-12}$ rad/s/$\sqrt{\text{Hz}}$, and aims to test the frame-dragging effect with a precision of 10% [12]. Besides, cold atom gyroscopes with such high precision will improve the precision of inertial navigation, especially when the Global Navigation Satellite System (GNSS) signal is unavailable or for deep space exploration [13]. Early studies have been carried out for interference experiments under microgravity platforms such as the drop tower [14,15], sounding rocket [16,17], parabolic flying plane [18,19], and the International Space Station [20-22]. However, in-orbit rotation measurement by AI has not been realized up to now.

There are several challenges to realizing the cold atom gyroscope in space. First, in microgravity and under a retroreflector Raman transition configuration, the energy levels of the Raman transitions of the two Raman laser pairs are degenerate. This will automatically form the double diffraction interference loop. How to use this interference configuration to measure the rotation is not implemented yet and needs to be verified. Second, in space, the rotation rate is usually much higher than the Earth's rotation rate. How to precisely extract the rotation in such a high dynamic condition without losing contrast still needs investigation. Third, the rotation-induced phase is related to the position and velocity of the atom cloud, so designing a proper scheme to eliminate this effect is a key problem in improving the rotation measurement precision.

In this article, we report the first rotation measurement result using a compact AI payload in the China space station (CSS). The point source interferometry (PSI) method [23,24] based on double diffraction Raman transition is realized under a microgravity environment. Spatial interference fringes are obtained and modulated using a built-in piezoelectric mirror, and rotation and acceleration are extracted from the interference fringes. This device is critical for precisely compensating the rotation rate of CSS, which is 15-fold higher than the earth's rotation rate. The complete expression of the rotation-induced phase, including the effects of the Raman laser's angles and the distributions of the cold atom cloud, is derived to measure the rotation. The ratio of the Raman laser's angles is optimized to eliminate the decoherent effect caused by the cold atom's position and velocity distribution, and these angles are self-calibrated in orbit to be better than 1 μrad by using the PSI fringes. Finally, we achieve a rotation measurement resolution of 50 μrad/s for a single shot and a long-term stability of 17 μrad/s for an average number of 32. After the systemic error correction, the measured rotation value is -1142±29 μrad/s, which agrees well with that measured by the classical gyroscope of the CSS platform. This work conducts the first AI-based gyroscope in space. The measurement scheme, the error estimation method, and the engineering design lay a foundation for the further development of cold atom gyroscopes in space.

## RESULTS
### Atom interference process of CSSAI

The China Space Station Atom Interferometer (CSSAI) is an integrated $^{85}$Rb-$^{87}$Rb dual-species AI payload. It has a size of 46 cm×33 cm×26 cm and a maximum power consumption of about 75 W. After the ground functional test, this payload was launched into the CSS in November 2022. The installation of the CSSAI in the CSS is shown in Fig. 1A. Section SI describes a brief outline of the payload and its operation. More details can be found in Ref. [25].

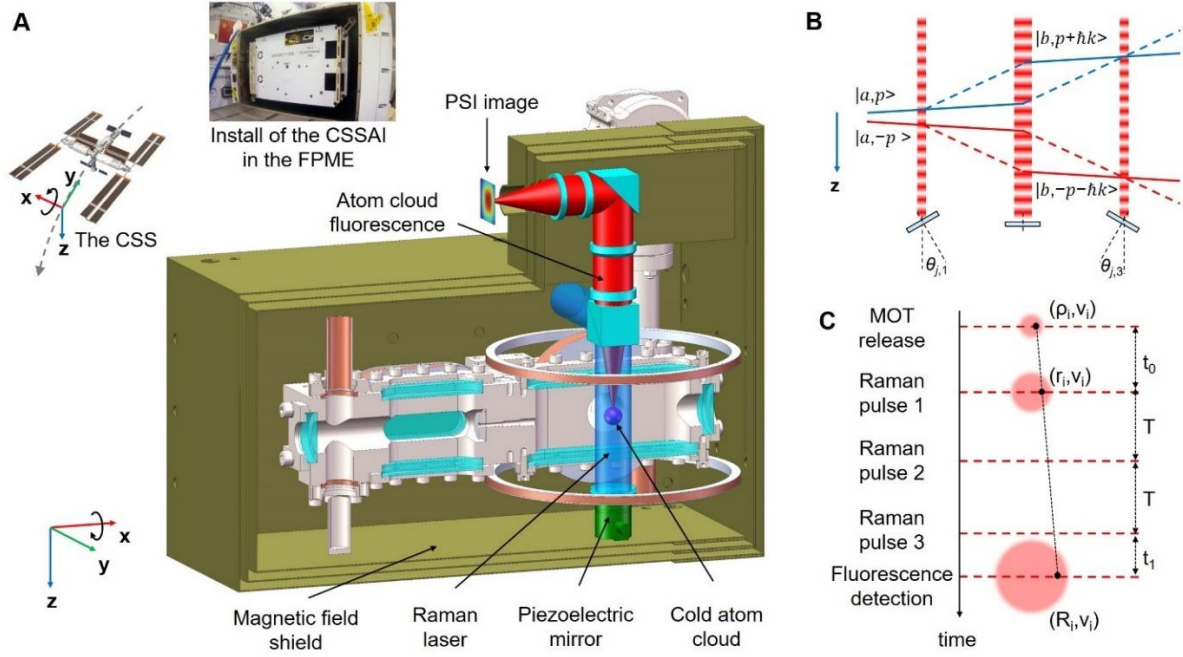

**Figure 1.** The working principle of the China Space Station Atom interferometer (CSSAI). (A) The China Space Station (CSS), the installed CSSAI in the Free-floating Platform for Microgravity Experiment (FPME), and the CSSAI's physical system profile. The cold atom cloud, the Raman laser for the point source interferometry (PSI), and the imaging of the fluorescence of the cold atom cloud are also illustrated in the physical system. (B) The double single diffraction (DSD) Raman transition and Raman interference loop scheme for the $^{87}$Rb atom. The state |a> and |b> represent the $|5^2S_{1/2}, F=1, M_F=0>$ and $|5^2S_{1/2}, F=2, M_F=0>$ states. (C) Atom position changes over time during the interference experiment.

The atom interference process of CSSAI is shown in Fig. 1B. $^{87}$Rb atom clouds with more than $10^8$ atoms with a temperature of 2 μK are produced using the two-dimensional magneto-optical trap (2D-MOT) and three-dimensional magneto-optical trap (3D-MOT). The atoms are then optically pumped from the $|5^2S_{1/2}, F=2>$ state to the $|5^2S_{1/2}, F=1>$ state. Because of the microgravity, the released cold atom cloud is at the center of the 3D-MOT chamber, and its velocity is zero. Linear polarization Raman laser is transmitted through a quarter wave plate and reflected by a mirror to produce the perpendicular linear polarization Raman laser pair. For the zero-velocity atom, the Raman laser pair will drive two Raman transitions in opposite directions, and the two-photon detuning of the Raman transitions will be the same. This will automatically induce the double diffraction Raman transition [19,26]. Here, we apply the double single diffraction (DSD) Raman transition [19] to create two symmetrical M-Z interference loops composed of $|5^2S_{1/2}, F=2, M_F=0\rangle$ and $|5^2S_{1/2}, F=1, M_F=0\rangle$ states, as shown in Fig. 1B. The duration of the Raman π pulse is set to be 17 μs, and the two-photon detuning is set to be 74 kHz to select two groups of cold atoms with opposite velocity. Because of the microgravity, the Doppler shift of the Raman transitions is constant, and the Raman laser's frequency chirp is not introduced. A closed-loop two-axis piezoelectric mirror is used to control the angle of the Raman laser during interference. This device is critical to inducing the PSI and creating the spatial interference fringe. After the interference, atoms in the $|5^2S_{1/2}, F=2\rangle$ state are fluorescently excited. The fluorescence passes through a polarization beam splitter and is imaged by a scientific CMOS camera in the same direction as the Raman laser, as shown in Fig. 1A. The imaging system has a magnification factor of 2.22±0.03, which is described in Section SII. To enhance the signal-to-noise ratio of the spatial interference fringe, the fluorescence image is averaged in an 1D curve, and a normalization method is designed to eliminate this background and normalize the spatial fringe's bias and amplitude fluctuation, which is introduced in Section SIII. Sine fitting is used to extract the fringe's phase and spatial frequency. The principal component analysis (PCA) method is also used to calculate the principal components for different orders.

**Derivation and optimization of the phase expression of PSI**

Considering the acceleration, rotation, and the Raman laser's angle, the spatial phase of the PSI [27] can be expressed as

$$\phi = k_{eff} a_z T^2 + \sum_i \delta_i k_{eff} [2\Omega_j v_i T^2 + \theta_{j,1} r_i + \theta_{j,3}(r_i + 2v_i T)], \tag{1}$$

where $k_{eff}$ is the effective wave vector, $T$ is the time interval between the Raman pulses, $i$ represents the $x$ and $y$ coordinate, and $j$ is the opposite of $i$, the symbol $\delta_i$ is defined as $\delta_x=1$ and $\delta_y=-1$, $a_z$ is the residual acceleration in the z-axes, $\Omega_j$ is the rotation of CSS in the $j$-axes and $\theta_{j,1}$ and $\theta_{j,3}$ are the angles of the Raman laser in the $j$-axes at the time of the 1st and 3rd Raman laser pulses relative to the angle at the time of the 2nd Raman laser pulse. $r_i$ and $v_i$ are the position and velocity of the atom at the time of the 1st Raman laser pulse. The imaging process will project the 3D population of the atom cloud to the 2D imaging plane. Because the phase variation and imaging plane are both in the x-y plane, the formula of the phase of the 2D spatial fringe is the same as that of Eq. (1).

The phase is related to $r_i$ and $v_i$. However, for the fluorescence image, what we measured is the spatial population of the atom at the fluorescence detection time, as shown in Fig. 1C. One has $R_i=r_i+v_i(2T+t_1)$, where $R_i$ is the atom's position at the fluorescence detection time and $t_1$ is the time interval between the 3rd Raman laser pulse and the fluorescence excitation laser pulse. For general cases, the position and velocity distribution of the atom will induce decoherence and period variation for the spatial fringe [28,29]. As simulated in Section IV and Fig. S6. If we submit $r_i=R_i-v_i(2T+t_1)$ in Eq. (1) and let the coefficient of $v_i$ to be zero. We will have the following relationship

$$\theta_{j,1} = \theta_{jo,1} = \frac{-t_1 \theta_{j,3} + 2\Omega_j T^2}{2T+t_1}. \tag{2}$$

We define $\theta_{jo,1}$ as the optimized angle. Then, $\phi$ is only related to $R_i$. Both the effects of the offset and distributions of the position and velocity of the atom are eliminated, and the contrast of the fringe will be maximized. The optimized phase has the following form.

$$\phi_o = k_{eff} a_z T^2 + \sum_{i=x,y} \delta_i f_{io} R_i, \tag{3}$$

$$f_{io} = \frac{2k_{eff}}{2T+t_1}(\theta_{j,3}T + \Omega_j T^2), \tag{4}$$

where $f_{io}$ is defined as the optimized spatial frequency of the spatial fringe.

For a more general case where Eq. (2) is not fulfilled, the exact expression for the phase and the spatial frequency is derived. By considering the atom cloud's position and velocity distributions and integrating the phase of Eq. (1) over them, strict analytical formulas for the phase and the spatial frequency are derived in the 'materials and methods' section as illustrated by Eq. (8) and (9). The exact solution of the spatial frequency is additionally related to the differential angle $\Delta\theta_j=\theta_{j,1}-\theta_{jo,1}$ and the width of the distributions of the atom cloud. So, by setting or measuring the various parameters in Eq. (9), and fitting the spatial frequency from the interference fringe, the rotation value can be calculated.

**In-orbit calibration of the Raman laser's angles**

The spatial frequency is related to the angle of the Raman laser, which is controlled by the piezoelectric mirror. The angle is proportional to the control voltage of the piezoelectric mirror with the relationship $\theta_j=\alpha_j V_j$, where $\alpha_j$ is defined as the voltage-angle coefficient. This coefficient is calibrated on the ground. However, the calibrated piezoelectric mirror is installed into the physical system and passes the mechanical and thermal tests. The coefficient might have some change. The accuracy angle is critical for calculating the rotation. However, one cannot calibrate it with external equipment after it is installed in the payload, especially after the payload is installed in the CSS.

Here, a self-calibration method is proposed and realized to measure the coefficient precisely by carrying out the in-orbit PSI experiment with T=50 ms. From Eq. (4), one can see that the optimized spatial frequency is proportional to the angle $\theta_{j,3}$, and thus proportional to $V_{j,3}$. So, if one sets a group value of $V_{j,3}$, adjusts the value of $V_{j,1}$ to satisfy the relationship of Eq. (2), and carries out the PSI experiment to measure the spatial frequency $f_{io}$, as shown in Fig. 2. The coefficient $\alpha_j$ could be obtained by linear fitting the values of $f_{io}$ and $V_{j,3}$. However, for an inaccurate value of $\alpha_j$, the calculated angles $\theta_j$ are inaccurate too, and Eq. (2) holds approximatively. The exact spatial frequency $f_i$ should be calculated by Eq. (9), where $f_i$ is not proportional to $V_{j,3}$ strictly.

To solve this problem, an iterative method is applied to get the exact value of $\alpha_j$. First, we set an initial set of estimated values of $\alpha_j$ and $\Omega_j$, calculate the differential angle $\Delta\theta_j$ by using Eq. (2), and calculate the differential spatial frequency $\Delta f_j$ by using Eq. (9). Then $f_i$ is subtracted by $\Delta f_j$ to obtain $f_{io}$. The values of $f_{io}$ and $V_{j,3}$ are linear fitted to get a new set of values of $\alpha_j$ and $\Omega_j$. This production continues until both $\alpha_j$ and $\Omega_j$ converge. The convergence process of $\Omega_j$ and $\alpha_j$ and the final convergent fitting curves for $f_{io}$ are shown in Fig. 2B, 2C, and 2A. The calculated coefficients are $\alpha_x$=116.75±0.41 μrad/V and $\alpha_y$=115.21±0.20 μrad/V respectively. From the obtained voltage-angle coefficients, the angle control precision is estimated to be 0.85 μrad and 0.33 μrad in the x and y directions. Besides the calibration of the Raman laser's angle. This produce also derives the value of the rotation rate of the CSS, which are $\Omega_x$=-1153±12 μrad/s and $\Omega_y$=-3.7±5.7μrad/s. The significant rotation rate in the x direction is due to the nadir-pointing rotation of the CSS around the earth.

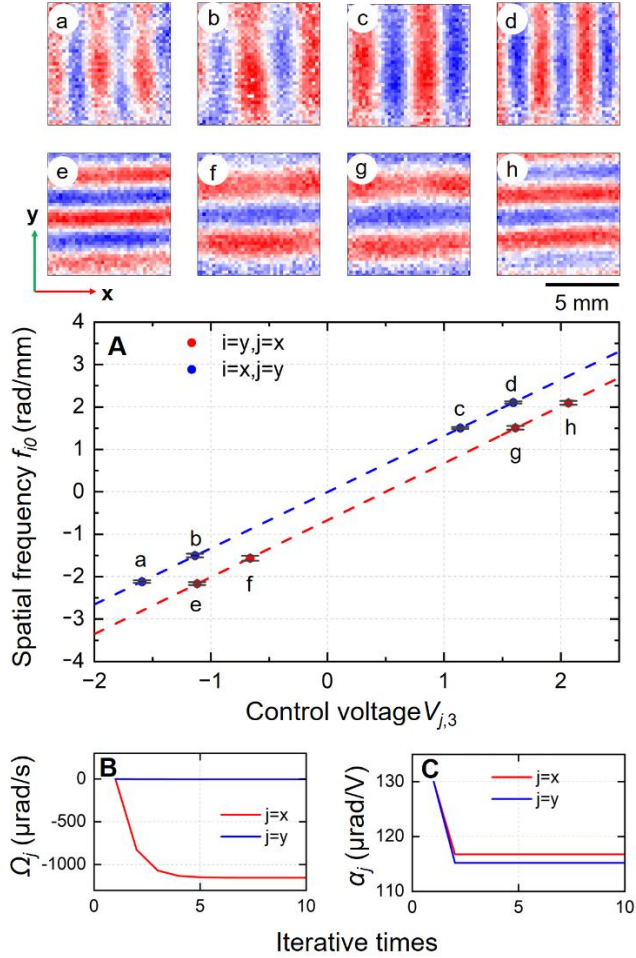

**Figure 2.** Calibration of the Raman laser's angle. (A) The first-order principal images of the 2D spatial interference fringes for the measurement points in Fig. 2B. (B) The relationship between the measured spatial frequencies $f_i$ and the control voltages of the piezoelectric mirror $V_{j,3}$. The dashed lines are the corresponding linear fitting curves. (C, D) The variation of the measured rotation values $\Omega_j$ and the voltage-angle coefficients $\alpha_j$ during the iterative process.

**Rotation extraction and error estimation**

To measure the rotation $\Omega_x$ and acceleration $a_z$ more precisely, PSI experiments with T=75 ms are carried out. The phase and spatial frequency are fitted from the fringes, and Eq. (8) and (9) are used to calculate the acceleration and rotation. To improve the measurement precision of rotation, the finite pulse width effect of the Raman laser is calculated and corrected (see 'Materials and methods' for detailed derivation). The calculated results of $\Omega_x$ and $a_z$ are shown in Fig. 3A and 3B. Due to the vibration of the CSS, the acceleration-induced phase variation exceeds $2\pi$, and the definite value for the acceleration is unknown. However, from the fitting residual phase, the measurement resolution of acceleration is estimated to be 1.0 μm/s² for a single shot. The rotation measurement resolution is 50 μrad/s for a single shot. The mean value is 1142 μrad/s, and the standard deviation is 101 μrad/s. The Allan deviation is shown in Fig. 3D, and the measurement resolution is 17 μrad/s for an averaged number of 32.

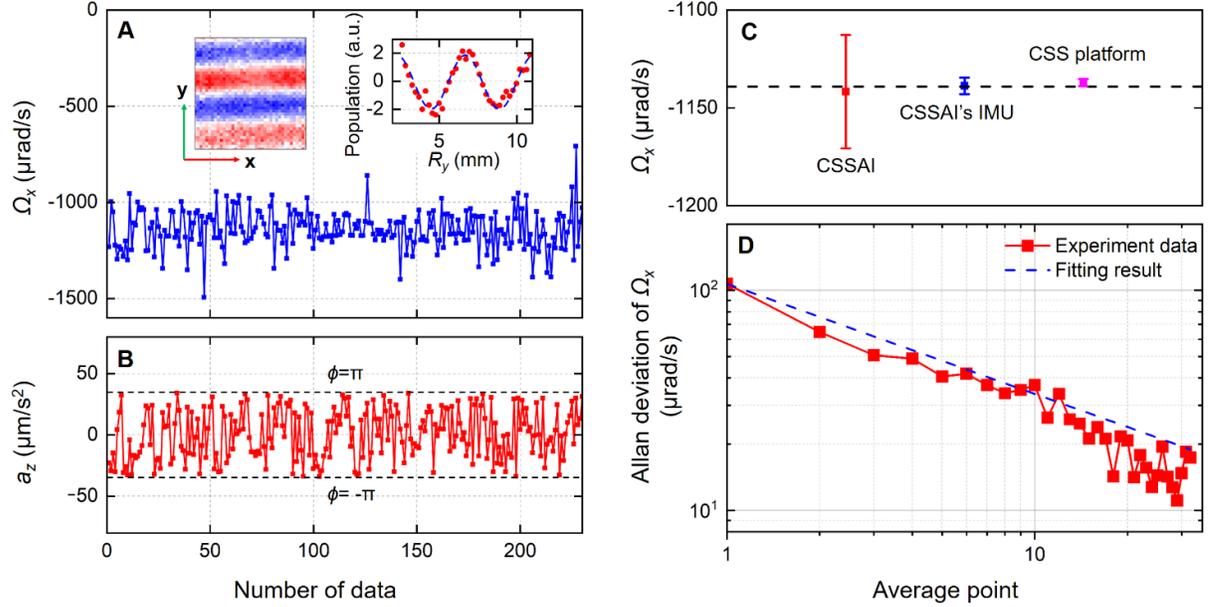

**Figure 3.** Rotation and acceleration measurement in space. (A) The measured value of the rotation $\Omega_x$. The inset images are the first-order principal images of the 2D spatial interference fringes for the measurement and a typical normalized 1D interference fringe. (B) The measured value of the acceleration $a_z$, the dashed lines represent the corresponding fitting phase of $\pi$ and $-\pi$. (C) Comparison of the measured rotation value by the CSSAI, CSSAI's inertial measuring unit (IMU), and the classical gyroscope of the CSS platform. The dashed line indicates their average value. (D)The Allan derivation of the measured rotation $\Omega_x$, the dashed line is its fitting curve with white noise.

Rotation measurement errors caused by the parameters' uncertainties are calculated according to Eq. (10) to estimate the measurement precision. Errors caused by the uncertainties of the magnification factor of the imaging system, the Raman laser's angle, the time sequence, the Raman laser frequency, and the distribution of the atom cloud are calculated in Section V.A-V.D. The magnetic field gradient-induced spatial frequency error is estimated by the ground measurement result in Section V.E. Other effects that only influence the phase but not the spatial frequency are not considered. These include the residual acceleration, the AC Stack shift, the multi-sideband effect [30,31], etc. The error terms are listed in Table 1. The largest error terms are the fitting noise of $f_i$, the uncertainty of the magnification factor, and the uncertainty of the Raman laser's angle. The measured rotation value is $\Omega_x$=-1142±29 μrad/s.

To check the measurement accuracy, we record the rotation measured by the CSSAI's inertial measuring unit (IMU) and the classical gyroscope of the CSS platform for the same time interval, which are -1138.7±4.1 μrad/s and -1137.0±1.8 μrad/s, respectively. These three measurement values are in good agreement, as shown in Fig. 3C.

**Table 1.** Error analysis for the rotation $\Omega_x$ with the PSI method in space.

| Parameters terms | Parameters values | Evaluated result (μrad/s) |
|---|---|---|
| Spatial frequency (fitting result) (rad/mm) | $f_y$=1.497±0.013 | -1142±17 |
| magnification factor of the imaging system (a.u.) | 2.22±0.03 | ±21 |
| Angles of 3rd Raman laser pulses (μrad) | $\theta_{x,3}$=202.94±0.72 | ±10 |
| Difference angle of $\theta_{x,1}$ (rad) | $\Delta\theta_x$=2.41±0.41 | ±1 |
| Interference time (μs) | $T$=75137.3±0.23 | ±3×10$^{-3}$ |
| Time before the 1st Raman pulse (μs) | $t_0$=43245.8±0.13 | ±2×10$^{-5}$ |
| Time after the 3rd Raman laser pulse (μs) | $t_1$=40146±10 | ±9×10$^{-2}$ |
| Width of the Raman π pulse (μs) | $\tau$=17±(5×10$^{-5}$) | ±6×10$^{-7}$ |
| Effective wave vector (m$^{-1}$) | $k_{eff}$=16105813.75±0.09 | ±9×10$^{-6}$ |
| Width of the MOT's position (mm) | $\sigma_{pi}$=0.427±0.013 | ±3×10$^{-2}$ |
| Width of the MOT's velocity (mm/s) | $\sigma_{vi}$=14.13±0.18 | ±1×10$^{-2}$ |
| Magnetic field | $B_0$=504.7±1.3 mG<br>$\gamma_{i,2}$=±1.3 G/m$^2$ | ±2×10$^{-1}$ |
| In total | | -1142±29 |

## DISCUSSION

This article introduces the integrated AI-based payload in the CSS and reports the first AI-based rotation measurement result in space. Spatial interference fringes are obtained using the PSI method based on the DSD interference scheme. The optimized ratio of the Raman laser's angle is derived to eliminate the decoherent effect caused by the cold atom cloud's position and velocity distribution. Formulas to extract the rotation are derived with the corrections of the offset of the Raman laser's angle and finite Raman laser pulse effect. In-orbit self-calibration of the Raman laser's angle is implemented to improve the accuracy of rotation measurement. In-orbit rotation measurement is carried out, and various errors are estimated. Real-time rotation measurement is achieved under a rotation rate that is 15-fold higher than the earth's rotation rate with a precision of 29 μrad/s.

The derived expressions for rotation measurement are adapted to the PSI experiment and could be applied to analyze the parameter requirement for more general cases. We take a Hyper-like experiment as an example [12]. The satellite is drag-free controlled and has a polar orbit with a height of 700 km. The frame-dragging effect-induced rotation is calculated to be oscillated with an amplitude of 2×10$^{-14}$ rad/s at twice the orbit frequency. a pair of atom gyroscopes with opposite atom velocities are installed in it. The interference time is set to be 10 s, the velocity of the atom cloud is set to be 0.2 m/s with a precision of 1 μm/s, double diffraction with 4 photon recoil is used to form the interference loop, the signal noise ratio of the interference fringe is set to be 10$^4$, and the data sampling rate is 1 Hz. Then, the measurement resolution of the rotation is calculated to be 4×10$^{-14}$ rad/s/$\sqrt{Hz}$, and a resolution of 7×10$^{-18}$ rad/s can be reached for 1 year of data integration. With these parameters, the experiment can measure the frame-dragging effect with a resolution of 0.04%. As illustrated in Eq. (1), the angle fluctuation of the Raman laser's mirrors will induce rotation measurement uncertainly. For a frame-dragging effect measurement precision of 0.1%, the requirement of the angle fluctuation of the mirrors at the signal frequency has to be less than $5\times10^{-11}$ rad. This put a strict constraint on the stability of the mechanical structure of the gyroscope.

The CSSAI can measure rotation in two dimensions and acceleration in one dimension. Increasing the number of Raman laser pairs allows this device to realize inertial measurement with complete vector components. By fusing the measurement data of the AI and the classical accelerometers and gyroscope by using the hybridization schemes [32,33], one can eliminate the deadtime effect and realize a space quantum inertial measurement unit, which can be used for inertial navigation of the spacecraft in orbit or deep space. The main constraints of the measurement precision of the rotation are the relatively short interference time and low effective atom velocity. These could be improved by preparing the atom cloud with ultra-cold temperatures using evaporative cooling and adiabatic cooling methods [14-21,34,35] and preparing the atom cloud with a reasonable non-zero velocity using the Bloch or Bragg coherent acceleration method [36,37].

The CSSAI also has the capability of measuring acceleration and differential acceleration. Potential applications cover earth gravity field measurement [38-40], equivalence principle (EP) tests [41,42], gravitational wave detection [43,44], dark matter detection [45,46], and test of general relativity effects [45]. The problem of large phase fluctuation caused by vibration could be solved by using hybridization measurement asocial with a classical accelerometer [32,33]. One can extract the in-orbit gravity acceleration by measuring the in-orbit residual acceleration with AI and combining it with the motion acceleration measured by the GNSS [47]. This could be used to invert or examine the gravity model of the Earth. The payload can measure the acceleration of the rubidium isotope synchronously, and the measured acceleration difference forms a quantum test of the EP in space [27,48,49]. Many noise and offset terms could be commonly rejected for the differential measurement, including the spacecraft's residual acceleration and the Raman laser's angle uncertainty, thus increasing the EP test precision.

## MATERIALS AND METHODS
**Derivation of the exact formulars of the phase and spatial frequency of PSI**

As illustrated in Fig. 1C. The position of the atom has the following relationship

$$\begin{aligned}R_i &= r_i + v_i(2T + t_1) \\ &= \rho_i + v_i t\end{aligned}, \quad (5)$$

where $\rho_i$ is the atom's position at the time of MOT release, $t=t_0+2T+t_1$ represents the total time, and $t_0$ is the time interval between the release of the MOT and the first Raman pulse. The distribution of the cold atom cloud at the time of MOT releasing is

$$F(\rho_i, v_i) = N_1 e^{-\frac{(\rho_i-\rho_{i0})^2}{2\sigma_{\rho i}^2}} e^{-\frac{(v_i-v_{i0})^2}{2\sigma_{vi}^2}}, \quad (6)$$

where $\rho_0$ and $\sigma_{\rho i}$ are the central and the distribution width of the position, $v_{i0}$ and $\sigma_{vi}$ are the central and the distribution width of the velocity, and $N_1$ is the normalization factor. Then, we substitute Eq. (5) into Eq. (1) and (6), and eliminate the variables $r_i$ and $\rho_i$. The population of the atom at the detection time can be calculated by integrating over $v_i$ with the following formula

$$P_I(R_i) = \int_{-\infty}^{\infty} P(R_i - v_i(2T + t_1), v_i) F(R_i - v_i t, v_i) dv_i. \quad (7)$$

We find the analytic expression of the integrated phase $\phi_I$ from the integrated population $P_I(R_i)$, which has the form

$$\begin{aligned}\phi_I = &\phi_o \\ &+ k_{eff} \sum_i \delta_i \left(\frac{t_0}{t} + \frac{t-t_0}{t}\frac{\sigma_{\rho i}^2}{\sigma_{vi}^2 t^2 + \sigma_{\rho i}^2}\right) R_i \Delta\theta_j \\ &+ k_{eff} \sum_i \delta_i \frac{t-t_0}{t} \cdot \frac{\sigma_{vi}^2 t^2 \rho_{i0} - \sigma_{\rho i}^2 v_{i0} t}{\sigma_{vi}^2 t^2 + \sigma_{\rho i}^2} \Delta\theta_j\end{aligned} \quad (8)$$

Besides the optimized phase $\phi_o$, the integrated phase is additionally related to the time parameters $t_0$ and $t$, the parameters of the cold atom cloud $\rho_{i0}$, $\sigma_{\rho i}$, $v_{i0}$, $\sigma_{vi}$, and the difference angle $\Delta\theta_j$. The spatial frequency of the integrated phase can be calculated as $f_i=\partial\phi_I/\partial R_i$, we define it the integrated spatial frequency, which has the following expression

$$\begin{aligned}f_i &= f_{io} + \Delta f_i \\ f_i &+ k_{eff}\left(\frac{t_0}{t} + \frac{t-t_0}{t}\frac{\sigma_{\rho i}^2}{\sigma_{vi}^2 t^2 + \sigma_{\rho i}^2}\right)\Delta\theta_j\end{aligned} \quad (9)$$

This is the exact formula for the PSI's spatial frequency. Besides the optimized spatial frequency $f_{i0}$, the integrated spatial frequency is additionally related to the differential angle $\Delta\theta_j$, and this formula is used to calculate the rotation $\Omega_j$ from the fitted value of $f_i$. The measurement uncertainty of $\Omega_j$ can be calculated by

$$d\Omega_j = \left(1/\frac{\partial f_i}{\partial \Omega_j}\right) df_i - \sum_k \left(\frac{\partial f_i}{\partial p_k}/\frac{\partial f_i}{\partial \Omega_j}\right) dp_k, \quad (10)$$

where $df_i$ is the fitting uncertainty of the spatial fringe, and $p_k$ and $dp_k$ represent the parameters and their uncertainties in the expression of $f_i$.

**Phase modification caused by the finite laser pulse effect**

The effect of the pulse width of the Raman laser is not considered in Eq. (1). This effect has to be considered for an accurate rotation measurement. The modified phase $\phi_m$ can be calculated by the sensitive function integrating method [50]. First, the time-dependent frequency and phase responses of the acceleration, rotation, and Raman laser's angle are derived, and then their corresponding sensitive functions are calculated. By multiplying these terms and integrating them over time, the expression of $\phi_m$ is as follows.

$$\begin{aligned}\phi_m = &k_{eff} a_z (T+\tau)\left(T + \frac{2\tau}{\pi}\right) \\ &+ k_{eff} \sum_i \delta_i 2\Omega_j v_i (T+\tau)\left(T + \frac{2\tau}{\pi}\right) \\ &+ k_{eff} \sum_i \delta_i \theta_{j,1}\left[r_i + \frac{(-2+\pi)v_i\tau}{2\pi}\right] \\ &+ k_{eff} \sum_i \delta_i \theta_{j,3}\left[r_i + v_i\frac{(4\pi T + 2\tau + 3\pi\tau)}{2\pi}\right]\end{aligned} \quad (11)$$

where $\tau$ is the width of the Raman $\pi$ pulse. By submit $r_i=R_i-v_i(2T+2\tau+t_1)$ in Eq. (11) and let the coefficient of $v_i$ to be zero, we find the optimized angle $\theta_{jmo,1}$ for the modified phase, which has the form

$$\theta_{jmo,1} = \frac{-(t_1 - \frac{\tau}{\pi} + \frac{\tau}{2})\theta_{j,3} + 2\Omega_j(T+\tau)\left(T+\frac{2\tau}{\pi}\right)}{2T+t_1+\frac{\tau}{\pi}+\frac{3\tau}{2}}. \quad (12)$$

Following the similar produce of the above method, one can derive the modified formulas of the integrated phase $\phi_I$ and integrated spatial frequency $f_i$ of $\phi_m$. these modified formulas are used to calibrate the Raman laser's angle and calculate the rotation. The formulas are not listed here because of their complex expressions.

**SUPPLEMENTARY DATA**
Supplementary data are available at NSR online.

**ACKNOWLEDGMENTS**
We would like to thank the support from the Technology and Engineering Center for Space Utilization, especially Hongen Zhong, Xuzhi Li, Shan Jin, and many others for their constructive discussions and technical support.

**FUNDING**
This work was supported by the second batch of the Scientific Experiment Project of the Space Application System of China Manned Space Program, the Space Application System of China Manned Space Program (JC2-0576), the Innovation Program for Quantum Science and Technology (2021ZD0300603, 2021ZD0300604), the Hubei Provincial Science and Technology Major Project (ZDZX2022000001), the China Postdoctoral Science Foundation (2020M672453), National Natural Science Foundation of China (12204493), and the Wuhan Dawn Plan Project (20230102010202082).

**AUTHOR CONTRIBUTIONS**
X. C. designed and charged the experiment. J. T. L., D. F. Z., and M. H. analyzed and processed the data. J. T. L., D. F. Z., W. Z. W., and Y. Z. operated the in-orbit experimental and collected the data. J. T. L. and J. F. conducted the ground comparison experiment. J. T. L., L. Z., C. H., and J. J. J. analyzed the precision of the rotation measurement. H. Y. S., Q. F. C., L. Q., X. L., Y. B. W., X. W. Z., R. B. L., and J. Q. Z. supported the in-orbit experiment in the optical system analysis, the electronic system analysis, and the software operation. M. Z. A., L. Z., S. Q. W., and Z. F. L. provided the required conditions of FPME for the in-orbit experiment. X. C. and J. T. L. prepared the manuscript. M. S. Z. and J. W. coordinated with the principal members as the project scientists. All authors have read and approved the final manuscript.

**REFERENCES**
1. Ciufolini I, Pavlis EC. A confirmation of the general relativistic prediction of the Lense–Thirring effect. *Nature* 2004; **431**: 958-960. 10.1038/nature03007
2. Everitt CWF, Adams M, Bencze W et al. Gravity Probe B data analysis status and potential for improved accuracy of scientific results. *Class Quantum Gravity* 2008; **25**: 114002. 10.1088/0264-9381/25/11/114002
3. Everitt CWF, DeBra DB, Parkinson BW et al. Gravity Probe B: Final Results of a Space Experiment to Test General Relativity. *Phys Rev Lett* 2011; **106**: 221101. 10.1103/PhysRevLett.106.221101
4. Ciufolini I, Paolozzi A, Pavlis EC et al. An improved test of the general relativistic effect of frame-dragging using the LARES and LAGEOS satellites. *Eur Phys J C* 2019; **79**: 872. 10.1140/epjc/s10052-019-7386-z

5. El-Sheimy N, Youssef A. Inertial sensors technologies for navigation applications: state of the art and future trends. *Satell Navig* 2020; **1**: 2. 10.1186/s43020-019-0001-5
6. Ciufolini I, Paolozzi A, Pavlis EC et al. The LARES 2 satellite, general relativity and fundamental physics. *Eur Phys J C* 2023; **83**: 87. 10.1140/epjc/s10052-023-11230-6
7. Di Virgilio ADV, Belfi J, Ni W-T et al. GINGER: A feasibility study. *The European Physical Journal Plus* 2017; **132**: 157. 10.1140/epjp/i2017-11452-6
8. Gustavson TL, Bouyer P, Kasevich MA. Precision Rotation Measurements with an Atom Interferometer Gyroscope. *Phys Rev Lett* 1997; **78**: 2046-2049. 10.1103/PhysRevLett.78.2046
9. Stockton JK, Takase K, Kasevich MA. Absolute Geodetic Rotation Measurement Using Atom Interferometry. *Phys Rev Lett* 2011; **107**: 133001. 10.1103/PhysRevLett.107.133001
10. Savoie D, Altorio M, Fang B et al. Interleaved atom interferometry for high-sensitivity inertial measurements. *Sci Adv* 2018; **4**: eaau7948. 10.1126/sciadv.aau7948
11. Yao Z-W, Chen H-H, Lu S-B et al. Self-alignment of a large-area dual-atom-interferometer gyroscope using parameter-decoupled phase-seeking calibrations. *Phys Rev A* 2021; **103**: 023319. 10.1103/PhysRevA.103.023319
12. Jentsch C, Müller T, Rasel EM et al. HYPER: A Satellite Mission in Fundamental Physics Based on High Precision Atom Interferometry. *Gen Relativ Gravit* 2004; **36**: 2197-2221. 10.1023/B:GERG.0000046179.26175.fa
13. Ning X, Gui M, Xu Y et al. INS/VNS/CNS integrated navigation method for planetary rovers. *Aerosp Sci Technol* 2016; **48**: 102-114. 10.1016/j.ast.2015.11.002
14. Rudolph J, Gaaloul N, Singh Y et al. Degenerate Quantum Gases in Microgravity. *Microgravity Sci Technol* 2011; **23**: 287-292. 10.1007/s12217-010-9247-0
15. Müntinga H, Ahlers H, Krutzik M et al. Interferometry with Bose-Einstein Condensates in Microgravity. *Phys Rev Lett* 2013; **110**: 093602. 10.1103/PhysRevLett.110.093602
16. Becker D, Lachmann MD, Seidel ST et al. Space-borne Bose–Einstein condensation for precision interferometry. *Nature* 2018; **562**: 391-395. 10.1038/s41586-018-0605-1
17. Lachmann MD, Ahlers H, Becker D et al. Ultracold atom interferometry in space. *Nat Commun* 2021; **12**: 1317. 10.1038/s41467-021-21628-z
18. Geiger R, Ménoret V, Stern G et al. Detecting inertial effects with airborne matter-wave interferometry. *Nat Commun* 2011; **2**: 474. 10.1038/ncomms1479
19. Barrett B, Antoni-Micollier L, Chichet L et al. Dual matter-wave inertial sensors in weightlessness. *Nat Commun* 2016; **7**: 13786. 10.1038/ncomms13786
20. Aveline DC, Williams JR, Elliott ER et al. Observation of Bose–Einstein condensates in an Earth-orbiting research lab. *Nature* 2020; **582**: 193-197. 10.1038/s41586-020-2346-1
21. Elliott ER, Aveline DC, Bigelow NP et al. Quantum gas mixtures and dual-species atom interferometry in space. *Nature* 2023; **623**: 502-508. 10.1038/s41586-023-06645-w
22. Williams JR, Sackett CA, Ahlers H et al. Pathfinder experiments with atom interferometry in the Cold Atom Lab onboard the International Space Station. *Nat Commun* 2024; **15**: 6414. 10.1038/s41467-024-50585-6
23. Dickerson SM, Hogan JM, Sugarbaker A et al. Multiaxis Inertial Sensing with Long-Time Point Source Atom Interferometry. *Phys Rev Lett* 2013; **111**: 083001. 10.1103/PhysRevLett.111.083001
24. Sugarbaker A, Dickerson SM, Hogan JM et al. Enhanced Atom Interferometer Readout through the Application of Phase Shear. *Phys Rev Lett* 2013; **111**: 113002. 10.1103/PhysRevLett.111.113002
25. He M, Chen X, Fang J et al. The space cold atom interferometer for testing the equivalence principle

in the China Space Station. *NPJ Microgravity* 2023; **9**: 58. 10.1038/s41526-023-00306-y
26. Lévèque T, Gauguet A, Michaud F et al. Enhancing the Area of a Raman Atom Interferometer Using a Versatile Double-Diffraction Technique. *Phys Rev Lett* 2009; **103**: 080405. 10.1103/PhysRevLett.103.080405
27. Asenbaum P, Overstreet C, Kim M et al. Atom-Interferometric Test of the Equivalence Principle at the $10^{-12}$ Level. *Phys Rev Lett* 2020; **125**: 191101. 10.1103/PhysRevLett.125.191101
28. Hoth GW, Pelle B, Riedl S et al. Point source atom interferometry with a cloud of finite size. *Appl Phys Lett* 2016; **109**: 071113. 10.1063/1.4961527
29. Chen Y-J, Hansen A, Hoth GW et al. Single-Source Multiaxis Cold-Atom Interferometer in a Centimeter-Scale Cell. *Phys Rev Appl* 2019; **12**: 014019. 10.1103/PhysRevApplied.12.014019
30. Peters A, Chung KY, Chu S. High-precision gravity measurements using atom interferometry. *Metrologia* 2001; **38**: 25. 10.1088/0026-1394/38/1/4
31. Carraz O, Charrière R, Cadoret M et al. Phase shift in an atom interferometer induced by the additional laser lines of a Raman laser generated by modulation. *Phys Rev A* 2012; **86**: 033605. 10.1103/PhysRevA.86.033605
32. Lautier J, Volodimer L, Hardin T et al. Hybridizing matter-wave and classical accelerometers. *Appl Phys Lett* 2014; **105**: 144102. 10.1063/1.4897358
33. Templier S, Cheiney P, d'Armagnac de Castanet Q et al. Tracking the vector acceleration with a hybrid quantum accelerometer triad. *Sci Adv* 2022; **8**: eadd3854. 10.1126/sciadv.add3854
34. Li L, Xiong W, Wang B et al. The Design, Realization, and Validation of the Scheme for Quantum Degenerate Research in Microgravity. *IEEE Photonics J* 2023; **15**: 1-8. 10.1109/JPHOT.2023.3266108
35. Li H, Wu B, Yu J et al. Momentum filtering scheme of cooling atomic clouds for the Chinese Space Station. *Chin Opt Lett* 2023; **21**: 080201. 10.3788/col202321.080201
36. Giltner DM, McGowan RW, Lee SA. Theoretical and experimental study of the Bragg scattering of atoms from a standing light wave. *Phys Rev A* 1995; **52**: 3966-3972. 10.1103/PhysRevA.52.3966
37. Cladé P, Guellati-Khélifa S, Nez F et al. Large Momentum Beam Splitter Using Bloch Oscillations. *Phys Rev Lett* 2009; **102**: 240402. 10.1103/PhysRevLett.102.240402
38. Chiow S-w, Williams J, Yu N. Laser-ranging long-baseline differential atom interferometers for space. *Phys Rev A* 2015; **92**: 063613. 10.1103/PhysRevA.92.063613
39. Douch K, Wu H, Schubert C et al. Simulation-based evaluation of a cold atom interferometry gradiometer concept for gravity field recovery. *Adv Space Res* 2018; **61**: 1307-1323. 10.1016/j.asr.2017.12.005
40. Lévèque T, Fallet C, Mandea M et al. Gravity field mapping using laser-coupled quantum accelerometers in space. *J Geodesy* 2021; **95**: 15. 10.1007/s00190-020-01462-9
41. Tino GM, Sorrentino F, Aguilera D et al. Precision Gravity Tests with Atom Interferometry in Space. *Nucl Phys B Proc Suppl* 2013; **243-244**: 203-217. 10.1016/j.nuclphysbps.2013.09.023
42. Williams J, Chiow S-w, Yu N et al. Quantum test of the equivalence principle and space-time aboard the International Space Station. *New J Phys* 2016; **18**: 025018. 10.1088/1367-2630/18/2/025018
43. Tino GM, Bassi A, Bianco G et al. SAGE: A proposal for a space atomic gravity explorer. *Eur Phys J D* 2019; **73**: 228. 10.1140/epjd/e2019-100324-6
44. Bertoldi A, Bongs K, Bouyer P et al. AEDGE: Atomic experiment for dark matter and gravity exploration in space. *Exp Astron* 2021; **51**: 1417-1426. 10.1007/s10686-021-09701-3
45. Amelino-Camelia G, Aplin K, Arndt M et al. GAUGE: the GrAnd Unification and Gravity Explorer.

*Exp Astron* 2009; **23**: 549-572. 10.1007/s10686-008-9086-9

46. Wolf P, Bordé CJ, Clairon A et al. Quantum physics exploring gravity in the outer solar system: the SAGAS project. *Exp Astron* 2009; **23**: 651-687. 10.1007/s10686-008-9118-5
47. Reigber C, Schwintzer P, Neumayer KH et al. The CHAMP-only earth gravity field model EIGEN-2. *Adv Space Res* 2003; **31**: 1883-1888. 10.1016/S0273-1177(03)00162-5
48. Zhou L, He C, Yan S-T et al. Joint mass-and-energy test of the equivalence principle at the $10^{-10}$ level using atoms with specified mass and internal energy. *Phys Rev A* 2021; **104**: 022822. 10.1103/PhysRevA.104.022822
49. Yuan L, Wu J, Yang S-J. Current Status and Prospects on High-Precision Quantum Tests of the Weak Equivalence Principle with Cold Atom Interferometry. *Symmetry* 2023, 10.3390/sym15091769
50. Le Gouët J, Cheinet P, Kim J et al. Influence of lasers propagation delay on the sensitivity of atom interferometers. *Eur Phys J D* 2007; **44**: 419-425. 10.1140/epjd/e2007-00218-2